\documentclass[preprint,showpacs,preprintnumbers,nofootinbib,amsmath,amssymb]{revtex4}

\usepackage{amssymb,color}
\usepackage{graphicx}
\usepackage{dcolumn}
\usepackage{bm}
\usepackage{subfigure}
\usepackage{hyperref}

\def\beqn{\begin{eqnarray}} 
\def\eeqn{\end{eqnarray}} 
\def\be{\begin{equation}}
\def\ee{\end{equation}}
\def\nn{\nonumber}

\def\mc{\mathcal}

\begin{document}

\title{Singlet Scalar Dark Matter: monochromatic gamma rays and metastable vacua}

\author{Stefano Profumo}
\email{profumo@scipp.ucsc.edu}
\affiliation{%
Santa Cruz Institute for Particle Physics and Department of Physics,\\ University of California, Santa Cruz CA 95064
}

\author{Lorenzo Ubaldi}
\email{ubaldi@physics.ucsc.edu}
\affiliation{%
Santa Cruz Institute for Particle Physics and Department of Physics,\\ University of California, Santa Cruz CA 95064
}

\author{Carroll Wainwright}
\email{cwainwri@ucsc.edu}
\affiliation{%
Santa Cruz Institute for Particle Physics and Department of Physics,\\ University of California, Santa Cruz CA 95064
}

\begin{abstract}
\noindent We calculate the pair-annihilation cross section of real scalar singlet dark matter into two mono-energetic photons. We derive constraints on the theory parameter space from the Fermi limits on gamma-ray lines, and we compare with current limits from direct dark matter detection. We show that the new limits, albeit typically relevant only when the dark matter mass is close to half the Standard Model Higgs mass, rule out regions of the theory parameter space that are otherwise not constrained by other observations or experiments. In particular, the new excluded regions partly overlap with the parameter space where real scalar singlet dark matter might explain the anomalous signals observed by CDMS. We also calculate the lifetime of unstable vacuum configurations in the scalar potential, and show that the gamma-ray limits are quite relevant in regions where the electro-weak vacuum is meta-stable with a lifetime longer than the age of the universe.
\end{abstract}




\maketitle

\section{Introduction}
Several possible approaches exist to embed ``{\em new phenomena}'' within ``{\em old paradigms}''. For instance, theoretical principles that apply to the old paradigm, or that are of great foundational significance, can be extended and used as guidelines to include new observed facts. A prime example of such principles are symmetries belonging to the old paradigm, or natural extensions of them. In some cases, this approach leads to including several unobserved components to the original theory, and hence redundancy in what is actually needed to interpret new observations. Another perfectly reasonable pathway is, instead, to pursue the idea of ``{\em minimality}'': what is the most economical extension to the pre-existing framework that allows to explain the new phenomena? While quantitatively defining the idea of a ``minimal'' extension is non-trivial, the particle content, or the number of needed new additional parameters are natural choices to quantify whether an extension to an elementary particle theory is or not economical.

One of the most compelling reasons to explore what might lie beyond the Standard Model (SM) of particle physics is the mysterious nature of the dark matter that dominates the matter content of the universe. In this context, theories that provide a dark matter candidate are widely considered more interesting extensions to the SM than those models that fall short of providing one. Supersymmetry \cite{susyrev} and theories with universal extra-dimensions \cite{uedrev} are examples of extensions of an old paradigm (here, the SM of particle physics) guided by symmetry principles that permit to explain new phenomena (the particle nature of the dark matter). The other pathway mentioned above, instead, has also been pursued successfully, postulating ad hoc, minimal extensions to the SM that encompass a dark matter particle candidate (for a systematic approach see e.g. Ref.~\cite{minimaldm}).

In many respects, what is widely considered to be the simplest, if not the most economical choice to embed a particle dark matter candidate into the framework of the SM, is to add a gauge-singlet real scalar field $S$ with renormalizable interactions only, and enforcing the $Z_2$ symmetry $S\to -S$. As we shall detail below, this theory, assuming $S$ is the only new degree of freedom at the electro-weak scale, only adds three new parameters to the SM: a mass term, a self-interaction term and a parameter that controls the coupling of the singlet to the other SM fields.

The theory we consider here has a quite long history, having been first envisioned by Veltman and Yndurain \cite{Veltman:1989vw}, who introduced a scalar ``$U$ particle'' to the SM and studied the impact of such particle on one-loop SM radiative corrections, in particular to $WW$ scattering. The theory was considered in a cosmological setup, and the scalar particle - there dubbed ``{\em scalar phantom}'' - as a dark matter candidate by Silveira and Zee in Ref.~\cite{Silveira:1985rk}. Most of the associated relevant phenomenology was worked out in Ref.~\cite{Silveira:1985rk}, including the calculation of the relic particle abundance from thermal freeze-out in the early universe, the scattering rate of the scalar particle off of baryons (direct detection), the effect on the SM Higgs decay and even the impact on the galactic cosmic-ray flux. Following that seminal work, a number of refined studies have considered the same, simple extension to the SM. In Ref.~\cite{McDonald:1993ex} the general case of an arbitrary number of {\em complex} singlet scalars was considered, with an emphasis again on cosmology in the early universe and direct detection. The specific case of one real singlet scalar was examined in great detail in Ref.~\cite{Burgess:2000yq}, including collider searches via anomalous Higgs decay patterns, dark matter self-interactions and constraints from the singlet potential.

The real scalar singlet extension to the SM was promoted in Ref.~\cite{Davoudiasl:2004be} to the status of ``New Minimal'' SM. With the advent of the Large Hadron Collider, several studies addressed the phenomenology of this paradigm with colliders, including e.g. Ref.~\cite{O'Connell:2006wi, Barger:2007im, Barger:2008jx}. A real scalar singlet also provides the possibility that the electro-weak phase transition be strongly first order, as needed to produce the observed baryon asymmetry in the context of electro-weak baryogenesis \cite{Pietroni:1992in, Profumo:2007wc}. In Ref.~\cite{Ponton:2008zv} and \cite{Kadastik:2009dj} TeV-scale scalar singlet extensions to the SM were shown to potentially have important implications for the recently observed cosmic-ray anomalies \cite{Adriani:2008zr, Abdo:2009zk, Grasso:2009ma}. 

Recent exciting results from direct dark matter experiments have triggered a renewed interest in real singlet scalar dark matter, that was invoked to interpret the DAMA \cite{Andreas:2008xy}, CDMS \cite{He:2008qm, Farina:2009ez, He:2009yd, Asano:2010yi,Kadastik:2009gx} and other anomalous signals like those detected with CoGeNT\cite{Andreas:2010dz}. The use of gamma-ray data, especially from the recently and successfully deployed Fermi Large Area Telescope (LAT), to detect a signature from singlet scalar dark matter has also been recently addressed in Ref.~\cite{Yaguna:2008hd, Barger:2010mc}. In the present study, we consider a channel that has not, to our knowledge, been addressed yet in this context: the pair annihilation into two, monochromatic high-energy gamma rays (see, however, Ref.~\cite{Badin:2009cf}). This channel is particularly relevant, given its unique spectral structure. This is unlike the previously considered continuum gamma-ray emission, a signal that could be confused with astrophysical backgrounds from e.g. emission from galactic cosmic rays or from milli-second pulsars. Also, the Fermi-LAT Collaboration recently assessed the observational constraints on searches for this peculiar spectral feature in Ref.~\cite{Abdo:2010nc}.

We find here that a particularly large and interesting region of the real singlet scalar dark matter parameter space for the two-photons annihilation channel is constrained by consideration of vacuum stability of the one-loop scalar potential, as recently studied in Ref.~\cite{Gonderinger:2009jp}. We reconsider here those constraints, in light of the possibility that even though the electro-weak minimum might be meta-stable, its lifetime for tunneling to the true minimum of the one-loop potential might be much longer than the age of the universe. If this is the case, the stability of the electro-weak vacuum is still valid on cosmological scales. We show in this study that constraints form the two-photon annihilation channel are particularly relevant in these regions of meta-stable electro-weak vacuum, that also overlap in some cases with parameter space portions compatible with the tentative positive direct dark matter detection signals reported by CDMS.

The outline of the present study is as follows: We first calculate, in section \ref{sec:cross}, the relevant $SS\to\gamma\gamma$ cross section. We then analyze the impact of the recent Fermi-LAT observations on the relevant parameter space in section \ref{sec:constraints}, and assess the impact for recent direct detection results in section \ref{sec:dirdet}. We explore the parameter space connected to meta-stable vacua in section \ref{sec:metastable}. Finally, we outline our conclusions and summarize our results in section \ref{sec:conclusions}.

\section{The $SS\to\gamma\gamma$ Cross section}\label{sec:cross}
We consider here a theory where a real scalar singlet $S$ is added to the particle content of the SM. Imposing the
$Z_2$ symmetry $S\to -S$ to the theory, so that the singlet is stable and becomes a candidate
for dark matter, the following Lagrangian exhausts all possible renormalizable terms (we follow here the notation of Ref.~\cite{Profumo:2007wc}):
\be
{\mc L} ={\mc L}_{\rm SM}+\frac{1}{2}\partial_\mu S \partial^\mu S-\frac{b_2}{2} S^2-\frac{b_4}{4} S^4-a_2 S^2
H^\dagger H
\ee
where ${\mc L}_{\rm SM}$ is the SM Lagrangian and $H$ is the SM Higgs doublet. This model adds the following three parameters to the SM: $b_2,b_4,a_2$. We require that, at the
minimum of the potential, the Higgs get a non-vanishing vacuum expectation value (VEV) $v=246$ GeV, but that the
singlet do not, $\langle S \rangle = 0$, in order to avoid making the singlet-like particle unstable, and thus not a viable dark matter candidate, as well as to prevent the appearance of problematic domain walls. In the unitary gauge, $H^\dagger = 1/\sqrt{2} (h+v,0)$ with $h$ real, after symmetry breaking, the scalar potential becomes
\be \label{eq:potential}
V(h,S)=-\frac{\mu^4}{4\lambda}-\mu^2 h^2+\lambda v h^3+\frac{\lambda}{4} h^4+\frac{1}{2}(b_2+a_2 v^2)S^2+\frac{b_4}{4} S^4+a_2 v S^2 h+\frac{a_2}{2}S^2 h^2,
\ee
where $\mu^2<0$, $\lambda$ is the quartic coupling for the Higgs, and $(-\mu^2/\lambda)^{1/2}=v$. This potential is bounded from below, at tree level, provided that $\lambda, b_4 \geq 0$, and $\lambda b_4\geq a^2_2$ for negative $a_2$. We see that the $S$ mass is
\be \label{eq:masses}
m^2_S = b_2 + a_2 v^2.
\ee

\begin{figure}[t]
\centering
\includegraphics[width=65 mm]{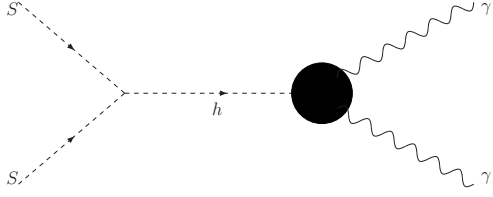}
\caption{Schematic Feynman diagram for the pair-annihilation of two singlets into two photons.}
\label{fig:diagram}
\end{figure}

As explained in the introduction, models of this kind have been studied in the literature, and constraints on the parameters have been derived, mostly from dark matter direct detection experiments. The aim of the present study is to put further constraints by studying the pair-annihilation channel into two photons, with $E_\gamma=m_S$, and by comparing with the 
photon lines limits obtained with Fermi-LAT  \cite{Abdo:2010nc}. In order to do so, we study the cross section for the process shown in Fig.~\ref{fig:diagram}.

The amplitude for the process reads
\be
\mc{M}_{SS\to \gamma \gamma} = 2 a_2 v \frac{i}{s-m_h^2-i\Gamma m_h}
\mc{M}_{h\to \gamma \gamma},
\ee
where $s$ is the center of mass energy squared, the total decay width $\Gamma=\Gamma(m_h)+\Gamma_S$, with $\Gamma(m_h)$ the decay width of the Higgs to SM particles and $\Gamma_S=\frac{a_2^2 v^2}{8\pi m_h} {\rm Re}\sqrt{1-4m^2_S/m^2_h}$ the decay width of the electro-weak Higgs to $SS$. The latter vanishes if the channel is kinematically forbidden, i.e. if $m_h<2m_S$. The annihilation cross section is given by
\be \label{eq:cross1}
\langle \sigma v\rangle_{\gamma \gamma} = \frac{1}{8 \pi s} |\mc{M}_{SS\to \gamma \gamma}|^2.
\ee
In the above Equations, $|\mc{M}_{h\to \gamma \gamma}|^2$ can be obtained from the result of the one-loop calculation of the width $\Gamma_{h\to \gamma \gamma}$ of the Higgs to two photons \cite{Shifman:1979eb, Howie}
\beqn \label{eq:Hwidth}
 \Gamma_{h\to \gamma \gamma} &=& \frac{1}{16 \pi m_h}|\mc{M}_{h\to \gamma \gamma}|^2 \nn\\
 &=&  \frac{\alpha^2 g^2}{1024 \pi^3 m_h}\frac{m_h^4}{M_W^2}|\sum_i N_{ci}e_i^2 F_i|^2.
\eeqn
In the annihilation process we study (see Fig.~\ref{fig:diagram}), the Higgs is, however, the exchanged particle and it can be off-shell. Therefore, in order to get the correct expression for $|\mc{M}_{h\to \gamma \gamma}|^2$ we need to substitute $m^2_h$ in Eq.~(\ref{eq:Hwidth}) with $s$. Thus we have
\be
|\mc{M}_{h\to \gamma \gamma}|^2 = \frac{\alpha^2 g^2}{64\pi^2}\frac{s^2}{M_W^2}|\sum_i N_{ci}e_i^2 F_i|^2,
\ee
where $i$ = spin-1/2 and spin-1 identifies the particle running in the loop, $N_{ci}$ is its color multiplicity, $e_i$ is the electric charge in units of $e$, and
\beqn \label{eq:Fs}
F_{1/2}&=& -2 \tau[1+(1-\tau)f(\tau)], \nn \\
F_1 &=& 2+3\tau+3\tau(2-\tau)f(\tau),
\eeqn
with $\tau = 4m^2_i/s$ and
\be
f(\tau)=\left\{ \begin{array}{ll}
\left[ \sin^{-1}(\sqrt{1/\tau})\right]^2, & {\rm if} \quad \tau\geq 1, \\
-\frac{1}{4}\left[\ln\left(\frac{1+\sqrt{1-\tau}}{1-\sqrt{1-\tau}}\right)-i \pi \right]^2, & {\rm if} \quad \tau < 1. 
\end{array} \right.
\ee
Plugging back into Eq.~(\ref{eq:cross1}) we have
\be \label{eq:finalcross}
\langle \sigma v\rangle_{\gamma \gamma}=a_2^2\frac{ \alpha^2}{32 \pi^3}\frac{s}{(s-m_h^2)^2+\Gamma^2 m_h^2}|\sum_i N_{ci}e_i^2 F_i|^2,
\ee
where we used $M_W=1/2 g v$. In the remainder of the paper we will consider the singlets to annihilate when they are non-relativistic, so that $s\simeq 4m_S^2$. 

\section{Constraints from $SS\to\gamma\gamma$ and Fermi-LAT observations}\label{sec:constraints}

\begin{figure}[t]
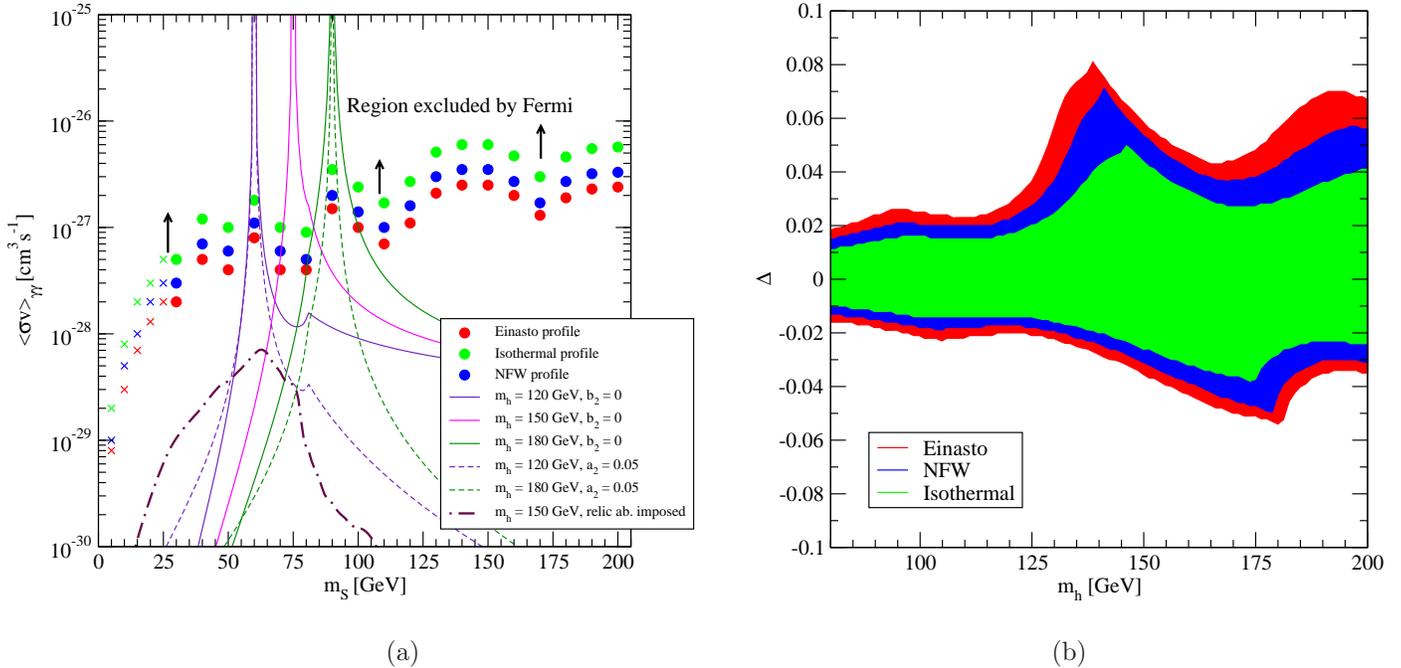

\centering
\mbox{\subfigure[]{
\hspace*{-1cm}\includegraphics[height=80 mm]{sigmam.eps} \label{fig:sigmam}}
\subfigure[]{
\includegraphics[height=80 mm]{Delta} \label{fig:delta}}}
\caption{{\em Left}: The pair-annihilation cross section of singlet scalar dark matter into two photons. Solid lines correspond to a Lagrangian with only $S^2 |H|^2$ interactions ($b_2=0$), and no relic abundance constraints, and a variety of values for the SU(2) Higgs mass $m_h=120,\ 150,\ 180$ GeV. Dashed lines correspond to a Lagrangian with $S^2 |H|^2$ interactions plus a mass term for $S$ ($b_2\neq 0$), with specified fixed values of $m_h$ and $a_2$, and again no relic abundance constraints. The dashed-dotted line features a quartic coupling, $a_2$, fixed to satisfy relic abundance constraint. The dots correspond to the limits from the Fermi-LAT collaboration \cite{Abdo:2009zk} for different dark matter profiles, whereas the crosses are an extrapolation of such limits to lower energies. {\em Right}: constraints from Fermi data on the plane defined by $m_h$ and $\Delta\equiv(m_S-m_h/2)/m_h$. Here we set $b_2=0$.}
\end{figure}

The Fermi-LAT Collaboration has recently searched for monochromatic $\gamma$ rays in the range 20-300 GeV\cite{Abdo:2009zk}, that would be produced by dark matter particle annihilation (for a study of the constraints from EGRET data see \cite{Pullen:2006sy}). We indicate the resulting limits on $\langle \sigma v\rangle_{\gamma \gamma}$  in Fig.~\ref{fig:sigmam} for three different representative dark matter density profiles: Einasto \cite{Einasto} (red dots), Isothermal \cite{Bahcall:1980fb}
 (green dots) and Navarro-Frenk-White (NFW) \cite{Navarro:1996gj}
 (blue dots) --- we refer the Reader to Ref.~\cite{Abdo:2009zk} for details on the analysis and on the specific assumptions for the dark matter density profiles. Given the recent interest in adopting the singlet scalar model to interpret signals that might be due to low mass dark matter ($m_S<30$ GeV), we thought it worth to extrapolate the Fermi limits to lower energies as well. In order to do so, and to be sufficiently conservative, we fixed the photon flux $\Phi$ to $5\times 10^{-9}$ cm$^{-2}$ s$^{-1}$, a value which is in line with those given for the lowest energies in Ref.~\cite{Abdo:2009zk}, and we used the fact that $\langle \sigma v\rangle_{\gamma \gamma} \propto m_S^2 \Phi$. The resulting limits on the cross section are shown in Fig.~\ref{fig:sigmam} for $m_S<30$ GeV as crosses, with the same color scheme for the three profiles. The different lines adopt different values for the mass of the electro-weak Higgs $m_h$ and for specific values of the parameters in the potential that remain unspecified after fixing $m_S$. In the solid lines we set $b_2=0$, thus corresponding to a Lagrangian with $S^2 |H|^2$ interactions plus a mass term for $S$ only, while for the dashed line we fix $a_2=0.05$. Finally, the dot-dashed line shows the cross section corresponding to $m_h=150$ GeV and the third parameter in the potential set to fulfill the requirement of an $S$ thermal relic abundance equal to the universal dark matter density.

It is clear from Fig.~\ref{fig:sigmam} that the cross section of Eq.~(\ref{eq:finalcross}) exceeds the Fermi limits only in a small region around the resonance, which happens for $m_S\simeq1/2 m_h$. This region is outlined with greater accuracy in Fig.~\ref{fig:delta}, where we plot $\Delta\equiv(m_S-m_h/2)/m_h$ versus $m_h$, after setting $b_2=0$. The shape of the Figure can be understood as a combination of two factors:
\begin{enumerate}
\item the non-trivial structure of the Fermi limits as a function of energy (i.e., here, as a function of the singlet mass) which depends on the astrophysical background and on the instrumental performance (e.g. point spread function and energy resolution);
\item the fact that, with $b_2=0$, we have $a_2=m_S^2/v^2$ and the annihilation cross section is proportional to $\frac{m_S^6}{(4m_S^2 - m_h^2)^2 + \Gamma_h^2 m_h^2}$. This second factor, in particular, explains the asymmetry of the plot.
\end{enumerate}

We point out that as we require $S$ to have a relic density consistent with WMAP, then $a_2$ is determined as a function of $m_S$ and $m_h$, and it becomes very small when $m_S=1/2 m_h$ (see {\em e.g.} \cite{Burgess:2000yq, He:2008qm}). This has the effect of canceling the resonance in our cross section, and, as a result we never exceed the Fermi constraints, as the dashed-dotted line in Fig.~\ref{fig:sigmam} shows.
  
We note that important constraints on the model under consideration here also stem from the continuum gamma-ray emission from $SS$ annihilation. In particular, one of the most stringent constraints comes from Fermi-LAT observations of local dwarf spheroidal galaxies \cite{Abdo:2010ex}. Although a systematic comparison with the present constraints from the monochromatic gamma-ray emission is beyond the scopes of the present analysis, it is useful to compare the two constraints in a few indicative cases. Let us consider, for instance, the green solid line ($m_h=180$ GeV, $b_2=0$), and $m_S=80$ GeV, a model which is right at the level of the $\gamma\gamma$ constraint. The corresponding total pair-annihilation cross section can be simply read off dividing by the Higgs decay branching fraction into two photons, which is around $10^{-3}$ for $m_h=2\times m_S=160$ GeV. We thus find that $\langle\sigma v\rangle_{\rm tot}\simeq{\rm few}\times 10^{-25}\ {\rm cm}^3{\rm s}^{-1}$. This is right around what found in Ref.~\cite{Abdo:2010ex} for a dark matter mass of 80 GeV (see fig.~3, upper-right corner). A similar comparison for other model-cases also indicates that the constraints we obtain here are comparable to those one would derive from Fermi-LAT observations of local dwarf spheroidal galaxies \cite{Abdo:2010ex}.

\begin{figure}[!h]
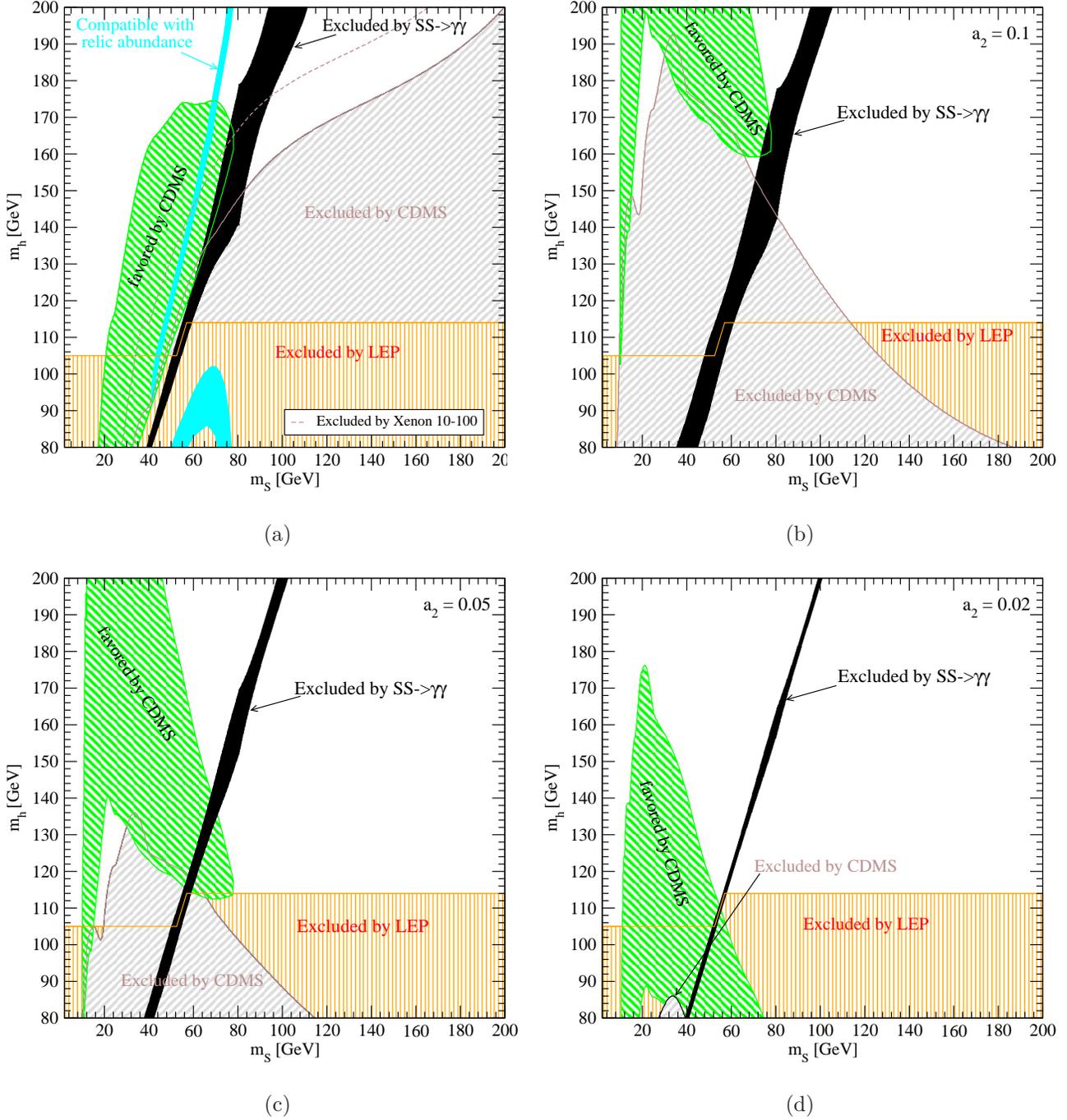

\centering
\mbox{\subfigure[]{
\hspace*{-0.5cm}\includegraphics[height=80 mm]{mhms_onlyquartic.eps} \label{fig:mhms_onlyquartic}}
\subfigure[]{
\includegraphics[height=80 mm]{mhms01.eps} \label{fig:mhms01}}}
\mbox{\subfigure[]{
\hspace*{-0.5cm}\includegraphics[height=80 mm]{mhms005.eps} \label{fig:mhms005}}
\subfigure[]{
\includegraphics[height=80 mm]{mhms002.eps} \label{fig:mhms002}}}
\caption{Top: Regions, on the $(m_S,m_h)$ parameter space, favored by the two events above background observed by CDMS at 78\% C.L. (green), and the regions excluded, at 90\% C.L., also by CDMS (grey); we indicate the region ruled out by LEP (orange), favored by the $S$ relic density (cyan), and excluded by Fermi searches for the monochromatic annihilation line, here coming from the process $SS\to\gamma\gamma$ (black). In the left panel we set $b_2=0$ and we also show the controversial limits from XENON \cite{Angle:2009xb, Aprile:2010um}, while in the right panel we have $b_2\neq 0$ and fix $a_2=0.1$. Bottom: As in Fig.~\ref{fig:mhms01}, but for $a_2=0.05$ (left) and 0.02 (right).}
\end{figure}

\clearpage

\section{Impact on direct detection results}\label{sec:dirdet}

In this section we explore how constraints on the singlet model from the annihilation $SS\to \gamma \gamma$ compare with the direct detection constraints. Only three parameters are relevant to our analysis: the Higgs mass $m_h$, $b_2$ (or, alternatively, $m_S$) and $a_2$\footnote{Note that the singlet self quartic coupling $b_4$ is completely irrelevant here.}. The direct detection constraints come from considering the spin-independent $S$-nucleon cross section \cite{Silveira:1985rk}
\be
\sigma_{\rm SI} = \frac{a_2^2 m^4_N f^2}{\pi m_S^2 m_h^4},
\ee
where $m_N$ is the nucleon mass, $f$ is the form factor. For numerical purposes, we set $f=1/3$, following here Ref.~\cite{Giedt:2009mr, Farina:2009ez}.

We will carry out the analysis for two different cases: (i) $b_2=0$ and (ii) $b_2\neq 0$.

\subsection{$b_2=0$.}
In this case, from Eq.~(\ref{eq:masses}) we have $a_2 = m_S^2/v^2$, and we are left with only two free parameters, that can be traded off for the two particle masses $m_h$ and $m_S$. The regions excluded by LEP \cite{LEP}, by CDMS \cite{Ahmed:2009zw} and by the Fermi results are shown in Fig.~\ref{fig:mhms_onlyquartic}. In the same plot, we also show the region compatible with the relic abundance and the one favored by CDMS at 78\% confidence level \cite{Ahmed:2009zw, Farina:2009ez}. Notice that while the excluded region is obtained from the 90\% C.L. upper limits on the WIMP-nucleon spin-independent cross section, the favored region comes from the two events observed by CDMS and is obtained at a different C.L. (namely, 78\%). We see that there is some overlap between the two regions, which should not be regarded as an inconsistency in our approach. Such an overlap would in fact change if we used different choices for the confident levels. We also indicate the controversial limits from XENON \cite{Angle:2009xb, Aprile:2010um} with a dashed brown line. We notice that the monochromatic photon line limits are competitive with respect to the direct detection limits for large Higgs masses and for $m_S\simeq m_h/2$. We also find a small portion of the parameter space compatible with the tentative signal observed by CDMS that is ruled out by the monochromatic photon lines limit. The line constraints, however, never overlap with the region where $S$ is thermally produced with the right relic abundance.

\subsection{$b_2\neq 0$.}
Without restrictions on $b_2$ we have to deal with a three-parameter space: $m_h, m_S, a_2$. We first consider the plane ($m_S,m_h$) and we show the excluded regions, as well as the CDMS favored region, for three different  values of $a_2$ in Fig.~\ref{fig:mhms01} (where we set $a_2=0.1$), \ref{fig:mhms005} ($a_2=0.05$) and \ref{fig:mhms002} ($a_2=0.02$). 

Decreasing the value of $a_2$ squeezes the width of the resonance, as a smaller value for $a_2$ reduces the cross section into two photons. At the same time, though, a reduced singlet-Higgs coupling also suppresses significantly the constraints from direct detection experiments. For $a_2\lesssim0.05$ and for $m_S\gtrsim 60$ GeV, the only constraints on the theory for viable values of the Higgs mass originate in fact from the two-photon annihilation mode.

\begin{figure}[t]
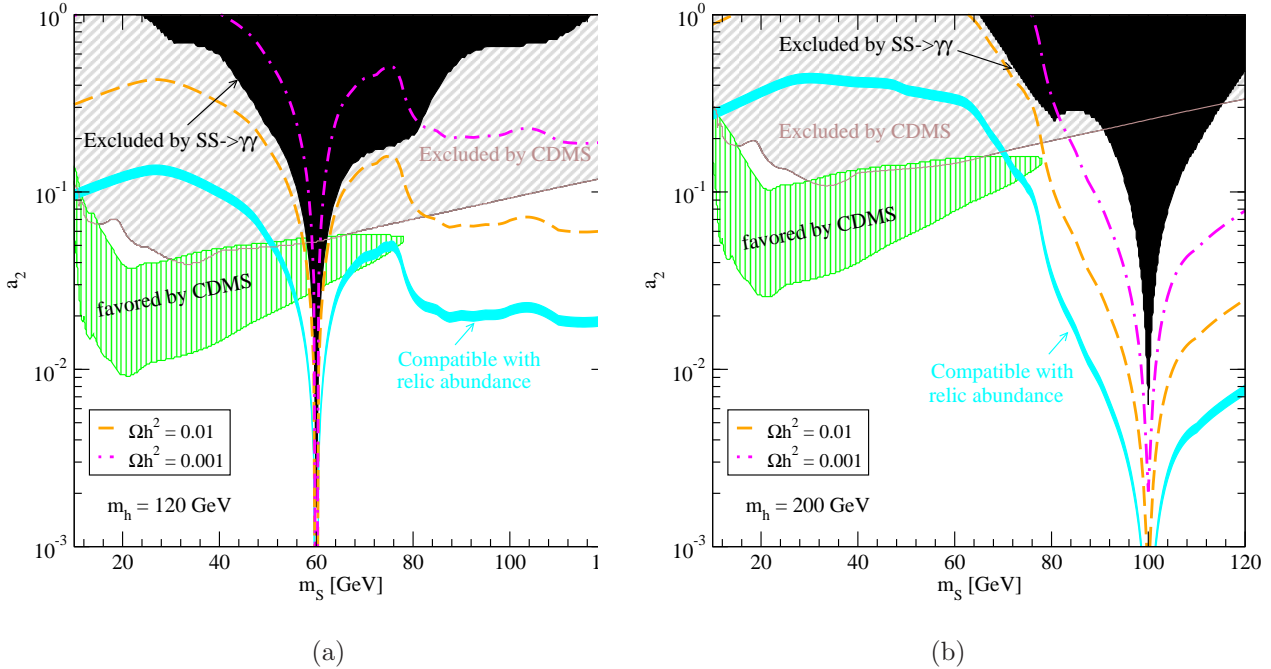

\centering
\mbox{\subfigure[]{
\hspace*{-0.5cm}\includegraphics[height=80 mm]{lams120new.eps}}
\subfigure[]{
\includegraphics[height=80 mm]{lams200new.eps}}}
\caption{Regions excluded (at 90\% C.L., brown) and regions marginally favored (at 78\% C.L., green) by the CDMS results and regions excluded by Fermi constraints (black) on the $SS\to\gamma\gamma$ process on the ($m_S,a_2$) parameter space. Within the cyan region the $S$ thermal relic abundance is compatible with the observed cosmological dark matter density; we also indicate the curves corresponding to a relic abundance of 0.01 (orange dashed lines) and of 0.001 (magenta dot-dashed lines). In the left panel we set $m_h=120$ GeV, while in the right panel $m_h=200$ GeV.\label{fig:lams}
}
\end{figure}

Lastly, we study the constraints on the plane ($m_S,a_2$) in Fig.~\ref{fig:lams}, where we fix $m_h=120$ GeV (a) and $m_h=200$ GeV (b). This cross-section of the theory parameter space illustrates that the $SS\to\gamma\gamma$ process excludes portions of the parameter space compatible with the CDMS putative signal (panel a), and it extends to values of $a_2$ smaller than those constrained by direct detection experiments, especially at growing values of the SM Higgs $m_h$ (panel b). We note that for $m_S>m_W$ three body final states produced by annihilation into $WW^*$  \cite{Yaguna:2010hn} as well as radiative electro-weak corrections \cite{Ciafaloni:2010ti} can also play a significant role.

\section{The role of the vacuum stability constraint}\label{sec:metastable}

The parameter space regions where the $SS\to\gamma\gamma$ annihilation mode puts further constraints upon the scalar singlet dark matter model is broadened by examining the stability of the Higgs vacuum. The tree-level potential given by Eq.~(\ref{eq:potential}) can in fact easily develop a second minimum in the singlet direction in addition to the expected minimum in the Higgs direction. This happens when the mass-squared term $b_2$ is negative and the coupling $a_2$ is large enough to overcome the negative mass-squared of the Higgs field. Specifically, the singlet will have a minimum at $S^2 = -b_2/b_4$ provided that $a_2 \cdot\frac{-b_2}{b_4}-2\mu^2 > 0$, which prevents the minimum from instead being a saddle point. If a second minimum exists and has a lower potential value than the minimum in the Higgs direction, then the physical vacuum state is unstable. However, there is a finite probability to tunnel to the new vacuum state. When the lifetime of the Higgs vacuum is longer than that of the universe, the vacuum is metastable and the theory is saved.
We perform here a stability analysis similar to Ref.~\cite{Gonderinger:2009jp}, but extend their study to allow for metastability and the possibility for one-loop corrections to lead to instabilities in the effective potential.

An important preliminary check is that the universe ends up in what will eventually evolve into the correct electro-weak minimum, at the electro-weak phase transition at high temperatures, instead of breaking the electro-weak symmetry in the singlet direction, and hence in what would then evolve to be the wrong zero-temperature vacuum. Finite temperature corrections to the effective potential tend to lift the potential away from the origin, and the lifting is strongest in regions with high mass particles. Even though we assume that the singlet scalar dark matter particle is relatively light in the electroweak phase, it could be quite heavy at the singlet minimum. In order to get a small dark matter mass either $a_2$ needs to be very small, in which case the electroweak minimum tends to be the true vacuum (see Fig.~\ref{fig:stability} below), or -$b_2$ needs to be very large, which gives rise to large masses in the singlet direction. Therefore, it is reasonable to expect that at temperatures near the electroweak phase transition the finite temperature potential lifts the singlet minimum above the electroweak minimum, and the universe gets stuck in the soon to be metastable electroweak phase. A detailed analysis of the finite-temperature vacuum structure of the theory lies however beyond the scope of this analysis.

The one loop corrections to the tree-level potential at zero temperature are
\begin{equation}
\label{eq:V1}
V_1(H^0,S) = \sum_i \pm \frac{n_i}{64\pi^2} m_i^4(H^0,S) \left[\log\left(\frac{m_i^2(H^0,S)}{\Lambda^2}\right) - c_i\right],
\end{equation}
where the sum is over all particle species, $n_i$ is the number of degrees of freedom per particle, $m_i$ is the field-dependent particle mass, $c_i =3/2$ for fermions and scalars and $5/6$ for gauge bosons, $\Lambda$ is the renormalization scale which we choose to be 1 TeV and  $H^0=h+v$ indicates the neutral real component of the SU(2) complex doublet $H$. The (tree-level) Higgs mass is $m_h^2 = \mu^2 + 3\lambda (H^0)^2 + a_2 S^2$, and the singlet mass is $m_S^2 = b_2+3b_4S^2+a_2(H^0)^2$. 
Reference~\cite{Gonderinger:2009jp} lists all other relevant field-dependent masses in its appendix. Note that they use slightly different notation than we do. They use $\lambda/6$, $m^2$ and $h$ where we use $\lambda$, $\mu^2$, and $H^0$. We follow the same procedure that they use for calculating the physical masses $M_h$ and $M_S$.

In order to ensure that the tree level minimum remains a minimum, we need to find the potential's second derivative:
\begin{equation}
\label{eq:d2V1}
\frac{\partial^2V_1}{\partial S^2} = \sum_i \frac{n_i}{32\pi^2} m_i^2(H^0,S) \left[\log\left(\frac{m_i^2(H^0,S)}{\Lambda^2}\right) - 1\right]  \frac{\partial^2 (m_i^2)}{\partial S^2}.
\end{equation}
Here, we took $c_i = 3/2$ and dropped terms containing ${\partial (m_i^2)}/{\partial S}$ which are zero at $S=0$. Unless the cutoff scale is taken to be smaller than the particle masses, the one-loop contribution tends to move the electroweak minimum towards instability. Usually this effect is not large enough to cancel the positive second derivative in the tree-level potential, but it can lead to instabilities in large sections of parameter space, as we will see below.

To examine the problem of metastability, we must calculate the tunneling rate per unit volume from the metastable to stable vacua. This rate has the form $\Gamma/V = Ae^{-S_E}$, where $S_E$ is the four-dimensional Euclidean action (see Ref.~\cite{coleman} for original work on the calculation of tunneling rates in field theory). The prefactor $A$ is generally difficult to calculate, but its exact value matters little in comparison to the rate's reliance upon $S_E$, so we can obtain an approximate solution on dimensional grounds. Assuming an $O(4)$ symmetry in the equations of motion, the action is
\begin{equation}
S_E = 2\pi^2 \int r^3dr\left[\frac{1}{2}\left(\frac{dH^0}{dr}\right)^2+\frac{1}{2}\left(\frac{dS}{dr}\right)^2 + V(H^0,S)\right]
\end{equation}
where $r$ is the Euclidean coordinate $r = (\vec{x},it)$. Minimizing $S_E$ produces the equations of motion
\begin{equation}
\label{eq:motion}
\frac{d^2H^0}{dr^2} + \frac{3}{r}\frac{dH^0}{dr} = \frac{\partial}{\partial H^0}V(H^0,S),
\end{equation}
and similarly for $H^0\rightarrow S$, with the boundary conditions
\begin{eqnarray}
\frac{dH^0}{dr} = \frac{dS}{dr} = 0 \text{ at } r = 0; \\
(H^0,S) = (v,0) \text{ at } r = \infty.
\end{eqnarray}
These equations describe a bubble of stable vacuum at $r=0$ embedded in a sea of the metastable electroweak vacuum.

If there were only one field, then Eq.~(\ref{eq:motion}) could easily be solved by the undershoot/overshoot method. One can exchange radius for time and then imagine the equation as describing a particle moving in the inverted potential $-V(H^0)$ with a peculiar `time'-dependent friction term $\frac{3}{r}(\frac{dH^0}{dr}+\frac{dS}{dr})$. The particle starts near the absolute maximum of $-V$ corresponding to the true vacuum, rolls down the potential, and then goes up again towards the maximum corresponding to the false vacuum. If the particle goes past the false vacuum, the initial conditions overshot the final conditions and they must be adjusted downward on the inverted potential. Conversely, if it does not make it to the false vacuum, the initial conditions undershot the final conditions and must be adjusted upwards.

The two dimensional case, however, is much more complicated. We can simplify it by assuming that tunneling occurs along a fixed path parametrized by its path length: $H^0=H^0(x)$, $S=S(x)$, and $(dH^0/dx)^2+(dS/dx)^2=1$. Equation~\ref{eq:motion} then simplifies to the one-dimensional case
\begin{equation}
\label{eq:motion2}
\frac{d^2x}{dr^2} + \frac{3}{r}\frac{dx}{dr} = \frac{\partial}{\partial x}V[H^0(x),S(x)],
\end{equation}
which we can solve by the undershoot/overshoot method. The trick then, is to choose the correct path. We do this by introducing a novel method of path deformation (see e.g. Ref.~\cite{konstandin2006} for another approach to finding the action. We will provide greater detail of our numerical algorithm in an upcoming paper.).

\begin{figure}[t] 
   \centering
   \includegraphics[width=6.5in]{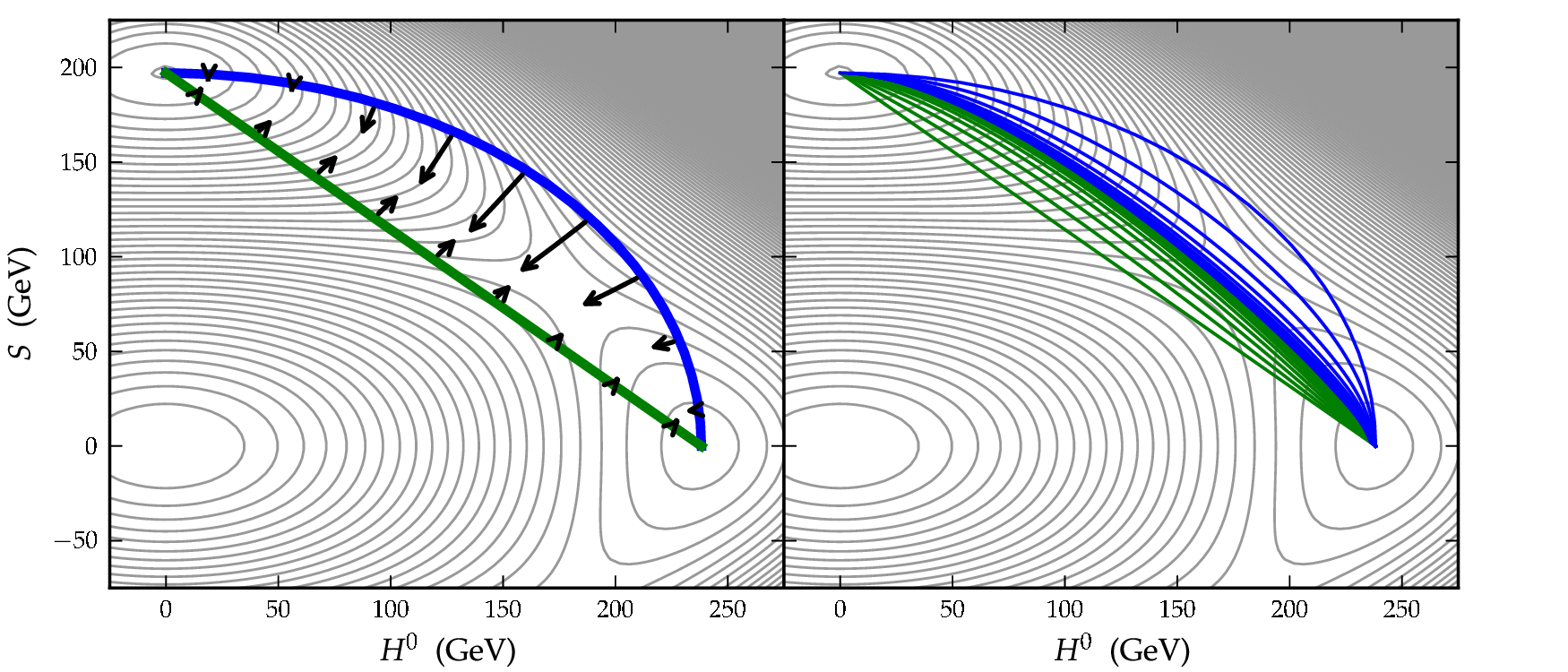} 
   \caption{Deformation of paths to solve the Euclidean equations of motion for $M_h = 120$ GeV, $M_S = 100$ GeV, and $a_2 = b_4 = 0.4$. The electroweak minimum is at $(H^0,S)=(246,0)$ GeV, and the stable singlet minimum is $(H^0,S)=(0,197)$ GeV. Left: We choose both a straight line and an elliptical path as first guesses for the direction of tunneling. Arrows denote the `normal' forces along each path. Right: The two guesses converge towards the correct solution. Each line represents 20 deformations with fixed step size. }
   \label{fig:deform}
\end{figure}

Let $\vec{\phi} = (H^0,S)$ describe the field coordinates. We can break the equations of motion into directions parallel and perpendicular to the direction of motion:
\begin{eqnarray}
\frac{d\vec{\phi}}{dx}\left(\frac{d^2x}{dr^2} + \frac{3}{r}\frac{dx}{dr}\right) = \left(\nabla_\phi V\right)_\parallel \\
\frac{d^2\vec{\phi}}{dx^2}\left(\frac{dx}{dr}\right)^2 = \left(\nabla_\phi V\right)_\perp.
\end{eqnarray}
Then, imagining $\vec{\phi}(x(r))$ as describing a particle moving on a fixed track, the quantity
\begin{equation}
N = \frac{d^2\vec{\phi}}{dx^2}\left(\frac{dx}{dr}\right)^2 - \left(\nabla_\phi V\right)_\perp
\end{equation} 
corresponds to the normal force exerted by the track upon the particle. If the track coincides with the natural direction of motion, the normal force will be zero. Otherwise the normal force will point in the direction of necessary path deformation (see Fig.~\ref{fig:deform}).

To execute the deformation, we first solve the one-dimensional equation of motion along a straight line between the two minima. We find $dx/dr$ at $n_{\rm points}=100$ evenly spaced points along the path, and then use this to find the normal force at those points. Each point deforms an amount $\Delta \vec{\phi} = \alpha L\vec{N}/|\nabla V|_{max}$, where $\alpha=0.002$ is our effective step size, $L$ is the length of the path, and $|\nabla V|_{max}$ is the maximum absolute gradient of the potential along the path. A more rigorous approach would be to use an adaptive step size, but a small constant step size is sufficient for our purposes. Typically, the deformation converges onto a new path in roughly 100 steps, at which point we re-solve the one-dimensional equation of motion. In our cases, we only need to repeat this process two or three times before we achieve an accuracy of about 1\% in the value of the Euclidean action. We check for convergence by picking a second starting path that lies on the other side of the final path (for example, an elliptical arc that connects the minima in the singlet and Higgs directions) and deforming from that direction.

All that is left is to approximate the pre-factor $A$ and find the critical value of $S_E$ for which we would have expected to see a phase transition. Here, we follow the argument in Ref.~\cite{sher1989}. Working in units of the electroweak scale, we set $A = 1$. The lifetime of the universe in electroweak units is $e^{101}$, and the fraction of the universe filled with stable phase as a function of time is $1-\exp(-\frac{\Gamma}{V}t^4)$ (see Ref.~\cite{Guth1981}). Therefore, in order for the Higgs vacuum to be metastable we require that the action $S_E$ be greater than 404.

\begin{figure}[t] 
   \centering
   \includegraphics[width=6.5in]{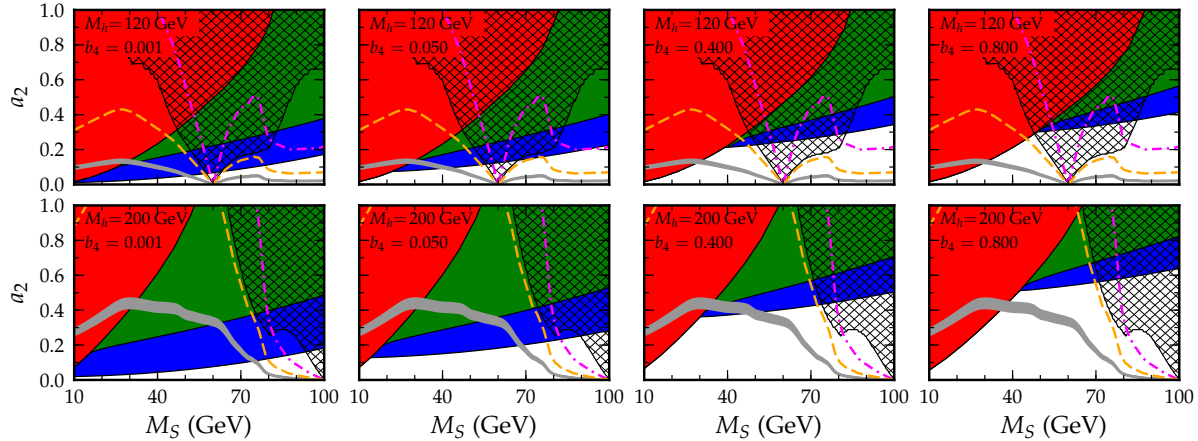} 
   \caption{Regions of stability for different parameters in the scalar singlet dark matter model. White regions are absolutely stable; that is, the minimum at $(H^0,S)=(v,0)$ is the true vacuum. Blue regions are metastable with lifetimes longer than that of the universe, while green regions are metastable with lifetimes shorter than that of the universe. In red regions, the field configuration corresponding to $(v,0)$ is not a minimum. We show the regions compatible with a thermal relic abundance with a grey band, and  points where the dark matter abundance equals 0.01 and 0.001 with dashed orange and dot-dashed magenta lines, respectively. Finally, the black hatched areas indicate regions excluded by the $SS \rightarrow \gamma\gamma$ constraint.}
   \label{fig:stability}
\end{figure}

We present our results in Fig.~\ref{fig:stability}. Ignoring the one-loop unstable region (red), we find identical areas of absolute stability as those in Ref.~\cite{Gonderinger:2009jp}. However, the total viable region of parameter space expands substantially by considering long-lived metastable vacua (blue regions).  The metastable states add roughly 0.1 and 0.2 to the maximum allowed $a_2$ value for low and high mass singlets, nearly doubling the allowed parameter space for theories with small singlet self-couplings ($b_4$). In the Figure, we show the regions compatible with a thermal relic abundance with a grey band, and  points where the dark matter abundance equals 0.01 and 0.001 with dashed orange and dot-dashed magenta lines, respectively. We notice, by inspecting Fig.~\ref{fig:lams} and comparing with Fig.~\ref{fig:stability} that specifically for small singlet self-coupling $b_4$, the region compatible with the putative CDMS signal interestingly overlaps largely with what we find are meta-stable but long-lived electro-weak vacuum configurations, for both heavy and light electro-weak masses.

We super-impose the constraints we obtain from Fermi observations and the calculation of the cross section for the annihilation process $SS\to\gamma\gamma$ we presented here (black hatched areas). We notice that, especially for low values of $b_4$, the two monochromatic photons annihilation mode sets strong constraints on the portion of the theory parameter space where the electro-weak vacuum is meta-stable with very long lifetimes. These regions also overlap with those compatible with the CDMS signal. Part of the meta-stable vacuum parameter space is nonetheless still open and not constrained by gamma-ray or direct detection experiments.

\section{Conclusions}\label{sec:conclusions}

We calculated the pair-annihilation cross section of a real, $Z_2$-symmetric scalar singlet extension of the SM into two photons, and we derived the constraints on the theory parameter space from the Fermi limits on monochromatic gamma-ray lines. We showed that this new class of constraints improve on limits from direct dark matter searches in certain regions of parameter space, especially where the singlet mass is close to half the SM Higgs mass. The limits we find rule out portions of the theory parameter space compatible with the tentative signal events found by the CDMS Collaboration. We also calculated the lifetime of unstable vacuum configurations in the scalar potential, and showed that the gamma-ray limits are quite relevant in regions where the electro-weak vacuum is meta-stable with a lifetime longer than the age of the universe. Those same regions also overlap with the portion of the theory parameter space compatible with the putative CDMS direct dark matter detection signal.

\section*{Acknowledgements}

We would like to thank Howard Haber and Michael Ramsey-Musolf for useful conversations. 
SP acknowledges support from the National Science Foundation, award PHY-0757911-001, and from an Outstanding Junior Investigator Award from the Department of Energy, DE-FG02-04ER41286. CW is supported by a National Science Foundation graduate Fellowship.


\begin{thebibliography}{99}

\bibitem{susyrev}
  G.~Jungman, M.~Kamionkowski, K.~Griest,
  Phys.\ Rept.\  {\bf 267}, 195-373 (1996).
  [hep-ph/9506380].

\bibitem{uedrev}
  D.~Hooper, S.~Profumo,
  Phys.\ Rept.\  {\bf 453}, 29-115 (2007).
  [hep-ph/0701197].

\bibitem{minimaldm}
  M.~Cirelli, N.~Fornengo, A.~Strumia,
  Nucl.\ Phys.\  {\bf B753}, 178-194 (2006).
  [hep-ph/0512090].

  \bibitem{Veltman:1989vw}
  M.~J.~G.~Veltman, F.~J.~Yndurain,
  Nucl.\ Phys.\  {\bf B325}, 1 (1989).
  
  \bibitem{Silveira:1985rk}
  V.~Silveira, A.~Zee,
  Phys.\ Lett.\  {\bf B161}, 136 (1985).
  
  \bibitem{McDonald:1993ex}
  J.~McDonald,
  Phys.\ Rev.\  {\bf D50}, 3637-3649 (1994).
  [hep-ph/0702143 [HEP-PH]].
  
  \bibitem{Burgess:2000yq}
  C.~P.~Burgess, M.~Pospelov, T.~ter Veldhuis,
  Nucl.\ Phys.\  {\bf B619}, 709-728 (2001).
  [hep-ph/0011335].
  
\bibitem{Davoudiasl:2004be}
  H.~Davoudiasl, R.~Kitano, T.~Li {\it et al.},
  Phys.\ Lett.\  {\bf B609}, 117-123 (2005).
  [hep-ph/0405097].

\bibitem{O'Connell:2006wi}
  D.~O'Connell, M.~J.~Ramsey-Musolf, M.~B.~Wise,
  Phys.\ Rev.\  {\bf D75}, 037701 (2007).
  [hep-ph/0611014].
  
  \bibitem{Barger:2007im}
  V.~Barger, P.~Langacker, M.~McCaskey {\it et al.},
  Phys.\ Rev.\  {\bf D77}, 035005 (2008).
  [arXiv:0706.4311 [hep-ph]].
  
  \bibitem{Barger:2008jx}
  V.~Barger, P.~Langacker, M.~McCaskey {\it et al.},
  Phys.\ Rev.\  {\bf D79}, 015018 (2009).
  [arXiv:0811.0393 [hep-ph]].
  
    \bibitem{Pietroni:1992in}
  M.~Pietroni,
  Nucl.\ Phys.\  {\bf B402}, 27-45 (1993).
  [hep-ph/9207227].
  
\bibitem{Profumo:2007wc}
  S.~Profumo, M.~J.~Ramsey-Musolf, G.~Shaughnessy,
  JHEP {\bf 0708}, 010 (2007).
  [arXiv:0705.2425 [hep-ph]].

  \bibitem{Ponton:2008zv}
  E.~Ponton, L.~Randall,
  JHEP {\bf 0904}, 080 (2009).
  [arXiv:0811.1029 [hep-ph]].
  
  \bibitem{Kadastik:2009dj}
  M.~Kadastik, K.~Kannike, M.~Raidal,
  Phys.\ Rev.\  {\bf D81}, 015002 (2010).
  [arXiv:0903.2475 [hep-ph]].
  
  \bibitem{Adriani:2008zr}
  O.~Adriani {\it et al.} [ PAMELA Collaboration ],
  Nature {\bf 458}, 607-609 (2009).
  [arXiv:0810.4995 [astro-ph]].

\bibitem{Abdo:2009zk}
  A.~A.~Abdo {\it et al.} [ The Fermi LAT Collaboration ],
  Phys.\ Rev.\ Lett.\  {\bf 102}, 181101 (2009).
  [arXiv:0905.0025 [astro-ph.HE]].
  
  \bibitem{Grasso:2009ma}
  D.~Grasso {\it et al.} [ FERMI-LAT Collaboration ],
  Astropart.\ Phys.\  {\bf 32}, 140-151 (2009).
  [arXiv:0905.0636 [astro-ph.HE]].
  
\bibitem{Andreas:2008xy}
  S.~Andreas, T.~Hambye, M.~H.~G.~Tytgat,
  JCAP {\bf 0810}, 034 (2008).
  [arXiv:0808.0255 [hep-ph]].

\bibitem{He:2008qm}
  X.~-G.~He, T.~Li, X.~-Q.~Li {\it et al.},
  Phys.\ Rev.\  {\bf D79}, 023521 (2009).
  [arXiv:0811.0658 [hep-ph]].

\bibitem{Farina:2009ez}
  M.~Farina, D.~Pappadopulo, A.~Strumia,
  Phys.\ Lett.\  {\bf B688}, 329-331 (2010).
  [arXiv:0912.5038 [hep-ph]].
  
\bibitem{He:2009yd}
  X.~-G.~He, T.~Li, X.~-Q.~Li {\it et al.},
  Phys.\ Lett.\  {\bf B688}, 332-336 (2010).
  [arXiv:0912.4722 [hep-ph]].

\bibitem{Asano:2010yi}
  M.~Asano and R.~Kitano,
  Phys.\ Rev.\  D {\bf 81} (2010) 054506
  [arXiv:1001.0486 [hep-ph]].

  \bibitem{Kadastik:2009gx}
  M.~Kadastik, K.~Kannike, A.~Racioppi {\it et al.},
  
  [arXiv:0912.3797 [hep-ph]].
  
\bibitem{Andreas:2010dz}
  S.~Andreas, C.~Arina, T.~Hambye {\it et al.},
  
  [arXiv:1003.2595 [hep-ph]].

\bibitem{Yaguna:2008hd}
  C.~E.~Yaguna,
  JCAP {\bf 0903}, 003 (2009).
  [arXiv:0810.4267 [hep-ph]].

\bibitem{Barger:2010mc}
  V.~Barger, Y.~Gao, M.~McCaskey {\it et al.},
  
  [arXiv:1008.1796 [hep-ph]].

\bibitem{Badin:2009cf}
  A.~Badin, G.~K.~Yeghiyan, A.~A.~Petrov,
  
  [arXiv:0909.5219 [hep-ph]].

  \bibitem{Abdo:2010nc}
  A.~A.~Abdo, M.~Ackermann, M.~Ajello {\it et al.},
  Phys.\ Rev.\ Lett.\  {\bf 104}, 091302 (2010).
  [arXiv:1001.4836 [astro-ph.HE]].

\bibitem{Gonderinger:2009jp}
 M.~Gonderinger, Y.~Li, H.~Patel and M.~J.~Ramsey-Musolf,
  JHEP {\bf 1001}, 053 (2010)
  [arXiv:0910.3167 [hep-ph]].

 
\bibitem{Shifman:1979eb}
  M.~A.~Shifman, A.~I.~Vainshtein, M.~B.~Voloshin {\it et al.},
  Sov.\ J.\ Nucl.\ Phys.\  {\bf 30}, 711-716 (1979).
  
  
 \bibitem{Howie}
  J.F.~Gunion, G.L.~Kane, H.E.~Haber and S.~Dawson , The Higgs hunter's guide. , Addison-Wesley, Reading, MA (1990).
  
  
\bibitem{Pullen:2006sy}
  A.~R.~Pullen, R.~-R.~Chary, M.~Kamionkowski,
  Phys.\ Rev.\  {\bf D76}, 063006 (2007).
  [astro-ph/0610295].

\bibitem{Einasto}
J.~Einasto, Trudy Inst. Astroz. Alma-Ata 51, 87 (1965).

\bibitem{Bahcall:1980fb}
  J.~N.~Bahcall, R.~M.~Soneira,
  Astrophys.\ J.\ Suppl.\  {\bf 44}, 73-110 (1980).

  
\bibitem{Navarro:1996gj}
  J.~F.~Navarro, C.~S.~Frenk, S.~D.~M.~White,
  Astrophys.\ J.\  {\bf 490}, 493-508 (1997).
  [astro-ph/9611107].


\bibitem{Abdo:2010ex}
  A.~A.~Abdo, M.~Ackermann, M.~Ajello {\it et al.},
  Astrophys.\ J.\  {\bf 712}, 147-158 (2010).
  [arXiv:1001.4531 [astro-ph.CO]].


\bibitem{Angle:2009xb}
  J.~Angle {\it et al.} [ XENON10 Collaboration ],
  Phys.\ Rev.\  {\bf D80}, 115005 (2009).
  [arXiv:0910.3698 [astro-ph.CO]].
  
  
\bibitem{Aprile:2010um}
  E.~Aprile {\it et al.} [ XENON100 Collaboration ],
  
  [arXiv:1005.0380 [astro-ph.CO]].

  
\bibitem{Giedt:2009mr}
  J.~Giedt, A.~W.~Thomas, R.~D.~Young,
  Phys.\ Rev.\ Lett.\  {\bf 103}, 201802 (2009).
  [arXiv:0907.4177 [hep-ph]].

  
  \bibitem{LEP}
  \url{http://lepewwg.web. 
cern.ch/LEPEWWG}
  
\bibitem{Ahmed:2009zw}
  Z.~Ahmed {\it et al.}  [The CDMS-II Collaboration],
  Science {\bf 327}, 1619 (2010)
  [arXiv:0912.3592 [astro-ph.CO]].


\bibitem{Yaguna:2010hn}
  C.~E.~Yaguna,
  Phys.\ Rev.\  {\bf D81}, 075024 (2010).
  [arXiv:1003.2730 [hep-ph]].
  
  \bibitem{Ciafaloni:2010ti}
  P.~Ciafaloni, D.~Comelli, A.~Riotto {\it et al.},
  
  [arXiv:1009.0224 [hep-ph]].

\bibitem{coleman}
S.~R.~Coleman,
  Phys.\ Rev.\  D {\bf 15}, 2929 (1977)
  [Erratum-ibid.\  D {\bf 16}, 1248 (1977)];
  C.~G.~.~Callan and S.~R.~Coleman,
  Phys.\ Rev.\  D {\bf 16}, 1762 (1977).

\bibitem{konstandin2006}
  T.~Konstandin and S.~J.~Huber,
  JCAP {\bf 0606}, 021 (2006).
  [hep-ph/0603081].


\bibitem{sher1989}
  M.~Sher,
  Phys.\ Rept.\  {\bf 179}, 273 (1989)

  \bibitem{Guth1981}
  A.~H.~Guth and E.~J.~Weinberg,
  Phys.\ Rev.\  D {\bf 23}, 876 (1981)


 \end{thebibliography}
\end{document}